\documentclass{elsart}

\usepackage{epsfig}
\usepackage{amssymb}

\begin{document}

\begin{frontmatter}

\title{Relativistic electromagnetic mass models in spherically symmetric spacetime}

\author[label1]{S.K. Maurya}\ead{sunil@unizwa.edu.om},
\author[label2]{Y.K. Gupta}\ead{kumar001947@gmail.com},
\author[label3]{Saibal Ray}\ead{saibal@iucaa.ernet.in}
\author[label4]{Vikram Chatterjee}\ead{vikphy1979@gmail.com}

\address[label1]{Department of Mathematical \& Physical
Sciences, College of Arts \& Science, University of Nizwa,
Nizwa-Sultanate of Oman}
\address[label2]{Department of Mathematics, Jaypee Institute
of Information Technology University, Sector-128 Noida (U.P.),
India}
\address[label3]{Department of Physics,
Government College of Engineering \& Ceramic Technology, Kolkata
700010, West Bengal, India}
\address[label4]{Department of Physics, Central Footwear
Training Centre, Kalipur, Budge Budge, South 24 Parganas 700138,
West Bengal, India}

\begin{abstract}
Under the static spherically symmetric Einstein-Maxwell spacetime
of embedding class one we explore possibility of electromagnetic
mass model where mass and other physical parameters have purely
electromagnetic origin~\cite{Tiwari1984,Gautreau1985,Gron1985}.
This work is in continuation of our earlier investigation
\cite{Maurya2015a} where we developed an algorithm and found out
three new solutions of electromagnetic mass models. In the present
letter we consider different metric potentials $\nu$ and $\lambda$
and analyzed them in a systematic way. It is observed that some of
the previous solutions related to electromagnetic mass models are
nothing but special cases of the presently obtained generalized
solution set. We further verify the solution set and show that
these are extremely applicable in the case of compact stars as
well as for understanding structure of the electron.
\end{abstract}


\end{frontmatter}

\section{Introduction}
The general theory of relativity (GTR), an outstanding extension
of the special theory of relativity with non-uniform reference
frame, was put forward by Einstein~\cite{OConnor1996} in the year
1915 - exactly 100 years ago! Till date this is considered as the
most profound and effective theory of gravitation. The field
theoretical effect of this geometric theory has been described by
Wheeler~\cite{Wheeler1990} in a poetic exposition as follows: {\it
``Matter tells space-time how to bent and space-time returns the
complement by telling matter how to move''.}

In the present context we employ GTR as our background canvas to
formulate solutions from {\it class 1 metric} and thereafter to
investigate {\it electromagnetic mass models}. As the present work
is a sequel of our earlier work~\cite{Maurya2015a} so we shall
refer this article of Maurya et al.~\cite{Maurya2015a} for
detailed discussions on class 1 metric as well as the
electromagnetic mass models. However, in the present work we
particularly give emphasis on the concept of electromagnetic mass
models for which a brief historical and philosophical review can
be obtained in the Ref.~\cite{Ray2007a}. On the other hand, for
the inclusion of charge in the spherical bodies one can look at
the Ref.~\cite{Maurya2015b}.

However, a special discussion on the electromagnetic mass models
seems required as provided by Gautreau~\cite{Gautreau1985}. Along
the line of thinking of Tiwari et al.~\cite{Tiwari1984} he shown
that the Einstein-Maxwell field equations of GTR can be used to
construct a Lorentzian model of an electron as an extended body
consisting of {\it pure charge and no
matter}~\cite{Lorentz1904,Wheeler1962,Feynman1964,Wilczek1999}.
However, in contrast with Lorentz's approach using inertial mass,
Gautreau~\cite{Gautreau1985} associated the mass of the electron
with its Schwarzschild gravitational mass and thus the field
equations for a Lorentz-type pure-charge extended electron could
be obtained by setting the matter terms equal to zero in the field
equations for a spherically symmetric charged perfect fluid. He
examined several explicit solutions to the pure-charge field
equations which we have shall use as a standard benchmark to
compare our solution set.

In connection to the above work on the electromagnetic mass models
it have been specially argued by Maurya et al.~\cite{Maurya2015a}
that most of the investigators
\cite{Cooperstock1978,Tiwari1984,Gautreau1985,Gron1985,Tiwari1986,Leon1987,Tiwari1991a,Tiwari1991b,Tiwari1991c,Tiwari1997,Ray2002,Ray2004,Ray2006,Ray2007b}
consider an {\it ad hoc} equation of state $\rho + p = 0$ (where
$\rho$ is the density and $p$ is the pressure), which suffers from
a negative pressure, and in the literature known as a 'false
vacuum' or 'degenerate vacuum' or
'$\rho$-vacuum'~\cite{Davies1984,Blome1984,Hogan1984,Kaiser1984}.
In the present investigation, however, for the construction of
electromagnetic mass models following Maurya et
al.~\cite{Maurya2015a} we also employ a different technique by
adopting an algorithm. We shall see later on that this  algorithm
will act as general platform to generate physically valid
solutions compatible with the spherically symmetric class one
metric.

The plan of the present work can be outlined as follows: we have
provided the static spherically symmetric spacetime and
Einstein-Maxwell field equations in the Sec. 2. In the Sec. 3 an
algorithm for class one metric has been developed from which we
construct a set of new general solutions. As a particular case
this solution set reduces to three sub-set as follows: (i) $a=0$
which corresponds to the charge analogue of the
Schwarzschild~\cite{Schwarzschild1916} interior solution, (ii)
$a=2b$ with $A=0$ which corresponds to the charge analogue of the
Kohlar-Chao~\cite{Kohler1965} interior solution, and (iii) $a=b$
which corresponds to the concept of electromagnetic mass model as
proposed by Lorentz~\cite{Lorentz1904} where $a$, $b$ and $A$ are
some constants. In the next Sec. 5 boundary conditions are
discussed to find out constants of integration. The Sec. 6 deals
with the solutions where critical analysis has been performed to
check several physical properties of the model whereas in the Sec.
7 we have particularly discussed some special features of the
models, firstly regarding validity with the stellar structure, and
secondly with the structure of the electron. We have made some
remarks in the concluding Sec. 8.

\section{The static spherically symmetric spacetime and Einstein-Maxwell field equations}

The Einstein-Maxwell field equations can be provided as usual
\begin{equation}
G_{j}^{i} =R_{j}^{i} -\frac{1}{2} R g_{j}^{i} =\kappa (T_{j}^{i}
+E_{j}^{i} ), \label{eq6}
\end{equation}
where $k = 8\pi$ is the Einstein constant ($G=c=1$, in the
relativistic units).

The matter distribution inside the star is assumed to be locally
perfect fluid and consequently $T_{j}^{i} $ and $E_{j}^{i}$, the
energy-momentum tensors for fluid distribution and electromagnetic
field respectively, are defined by
\begin{equation}
T_{j}^{i} =[\rho +p)v^{i} v_{j} -p\delta _{j}^{i} ], \label{eq7}
\end{equation}

\begin{equation}
E_{j}^{i} =\frac{1}{4\pi} (-F^{im} F_{jm} +\frac{1}{4} \delta
_{j}^{i} F^{m\, n} F_{m\, n} ), \label{eq8}
\end{equation}
where $v^{i}$ is the four-velocity as $e^{-\nu(r)/2} v^{i} =
\delta _{4}^{i} $, $\rho$ is the energy density and $p$ is the
fluid pressure of the matter distribution.

Now the anti-symmetric electromagnetic field tensor, $F_{ij}$,
satisfies the Maxwell equations
\begin{equation}
F_{ik,j} +F_{kj,i} +F_{ji,k} =0, \label{eq10}
\end{equation}

\begin{equation}
\frac{\partial }{\partial x^{k} } (\sqrt{-g} F^{ik} )=-4\pi
\sqrt{-g} j^{i}, \label{eq11}
\end{equation}
where $g$ is the determinant of quantities $g_{ij}$ in
Eq.~(\ref{eq1}) and is defined by $g = - e^{(\nu+\lambda)}r^4
sin^2\theta $.

The only non-vanishing components of electromagnetic field tensor
are $F^{41}$ and $F_{14} $ which describe the radial component of
the electric field and are related as $F^{41} = -F^{14}$. From Eq.
(\ref{eq11}), we can obtain the following expression for the
electric field
\begin{equation}
F^{41} =e^{-\frac{(\lambda +\nu )}{2} } \frac{q(r)}{r^{2} } \label{eq12}
\end{equation}
where $q(r)$ represents the total charge contained within the
sphere of radius $r$ and is defined by
\begin{equation}
q(r) = r^{2} \sqrt{-F_{14} F^{14} } = r^{2} F^{41} e^{(\lambda
+\nu )/2} = r^2 E = 4\pi \int_{0}^{r}\sigma \, r^{2} e^{\lambda
/2} dr, \label{eq13}
\end{equation}
where $\sigma $ is the charge density.

Now, following the work of Maurya et al. \cite{Maurya2015a} here
we consider the static spherically symmetric metric in the form
\begin{equation}
ds^{2} =-e^{\lambda } dr^{2} -r^{2} (d\theta ^{2} +\sin ^{2}
\theta d\varphi ^{2} )+e^{\nu } dt^{2}. \label{eq1}
\end{equation}
The above metric represents spacetime of embedding class one if it
satisfies the Karmarkar condition \cite{Karmarkar1948}
\begin{equation}
R_{1414} =\frac{R_{1212} R_{3434} + R_{1224} R_{1334} }{R_{2323} }
\label{eq2}
\end{equation}
along with the constraint $R_{2323} \neq 0 $ \cite{Pandey1982}.

Therefore, by the application of above condition in Eq.
(\ref{eq1}) we obtain the following second order differential
equation
\begin{equation}
\frac{\lambda' \nu'}{(1-e^{\lambda } )}=-2(\nu'' +\nu^{\prime 2}
)+ \nu^{\prime 2} +\lambda' \nu'~; \quad e^{\lambda } \neq 1,
\label{eq3}
\end{equation}
where $\nu(r)$ and $\lambda(r)$ are metric potentials and depends
only on the radial coordinate $r$.

After manipulation, the solution of the second order differential
equation~(\ref{eq3}), can be obtained as
\begin{equation}
e^{\lambda } =\left( 1+K\frac{\nu ^{\prime2} e^{\nu } }{4}
\right). \label{eq4}
\end{equation}
Here $K$ is a non-zero arbitrary constant, $\nu'(r)\neq 0$,
$e^{\lambda (0)} =1$ and $\nu'(0)=0$.

For the above spherically symmetric metric (\ref{eq1}), the
Einstein-Maxwell field equations (\ref{eq6}) can be expressed as
the following set of ordinary differential
equations~\cite{Maurya2015a,Maurya2015b}
\begin{equation}
\frac{\nu'}{r} e^{-\lambda } -\frac{(1-e^{-\lambda } )}{r^{2} }
=-\kappa T_{1}^{1}=\kappa p-\frac{q}{r^{4} }^{2}, \label{eq15}
\end{equation}

\begin{equation}
-\kappa \, T_{3}^{3} =\left[ \frac{\nu''}{2}
-\frac{\lambda'\nu'}{4} +\frac{\nu'^{2} }{4}
+\frac{\nu'-\lambda'}{2r} \right] e^{-\lambda } =-\kappa T_{2}^{2}
=\kappa p+\frac{q}{r^{4} }^{2},\label{eq16}
\end{equation}

\begin{equation}
\frac{\lambda'}{r} e^{-\lambda } +\frac{(1-e^{-\lambda } )}{r^{2}
} =\kappa T_{4}^{4} =\kappa \rho +\frac{q}{r^{4} }^{2},
\label{eq17}
\end{equation}
where the prime denotes differentiation with respect to $r$.

Therefore, by incorporating Eq.~(\ref{eq4}) in the set of
Eqs.~(\ref{eq15}) - (\ref{eq17}), we get
\begin{equation}
\frac{\nu ' }{r^{2} (4+K\nu ^{\prime2} e^{\nu } )} (4r-K\nu '
)=\kappa p-\frac{q^{2} }{r^{4} }, \label{eq18}
\end{equation}

\begin{equation}
\frac{4}{(4+K\nu ^{\prime2} e^{\nu } )} \left( \frac{\nu ' }{2r}
-\frac{(K\nu ' e^{\nu } -2r) (2\nu '' +\nu ^{\prime2} )}{2r(4+K\nu
^{\prime2} e^{\nu } )} \right)=\kappa p+\frac{q^{2} }{r^{4} },
\label{eq19}
\end{equation}

\begin{equation}
\frac{Ke^{\nu } \nu ' }{(4+K\nu ^{\prime2} e^{\nu } )} \left(
\frac{4(2\nu '' +\nu ^{\prime2} )}{(4+K\nu ^{\prime2} e^{\nu } )}
+\frac{\nu ' }{r} \right) =\kappa \rho +\frac{q^{2} }{r^{4} }.
\label{eq20}
\end{equation}

Along with these Eqs. ~(\ref{eq15}) - (\ref{eq17}), we also
include the pressure isotropy condition and pressure gradient as
follows
\begin{equation}
\left( \frac{K\nu ' e^{\nu }}{2r} -1\right) \left( \frac{2\nu '
}{r (4+K\nu ^{\prime2} e^{\nu } )} -\frac{4(2\nu '' +\nu
^{\prime2} )}{(4+K\nu ^{\prime2} e^{\nu } )^{2} } \right)
=\frac{2q^{2} }{r^{4} }, \label{eq21}
\end{equation}

\begin{equation}
\frac{dp}{dr} =-\frac{M_{G} (r)(\rho +p)}{r^{2} } e^{(\lambda -\nu
)/2} +\frac{q}{4 \pi r^{4} } \frac{dq}{dr}, \label{eq32}
\end{equation}
where $M_{G}$ is the gravitational mass within the radius $r$ and
is given by
\begin{equation}
M_{G} (r)=\frac{1}{2} r^{2}\nu ' e^{(\nu -\lambda )/2}.
\label{eq33}
\end{equation}

The above Eq.~(\ref{eq32}) represents the charged generalization
of the Tolman-Oppenheimer-Volkoff
(TOV)~\cite{Tolman1939,Oppenheimer1939} equation of hydrostatic
equilibrium or equation of continuity.

It have been argued by \cite{Maurya2015a} that if charge vanishes
in a charged fluid of embedding class one then survived neutral
counterpart will only be either the Schwarzschild
~\cite{Schwarzschild1916} interior solution (or its special cases
de-sitter universe or Einstein's universe) or
Kohler-Chao~\cite{Kohler1965} solution otherwise either charge
cannot be zero or the survived space-time metric is flat.

\section{The algorithm of electromagnetic mass models for class one metric}

It can be shown that in the presence of electrical charge the
fluid sphere under consideration can be defined by the following
metric functions:
\begin{equation}
e^{-\lambda} = 1-\frac{2m(r)}{r} +\frac{q^{2} }{r^{2} },
\label{eq28}
\end{equation}

\begin{equation}
\nu' =\frac{\left( \kappa r p+\frac{2m}{r^{2} } -\frac{2 q^{2}
}{r^{3} } \right) }{\left( 1-\frac{2m}{r} +\frac{q^{2} }{r^{2} }
\right) }, \label{eq31}
\end{equation}
where $m(r)$ is the mass function which in the explicit form can
be written as \cite{Florides1983}
\begin{equation}
m(r)=\frac{\kappa }{2} \int\rho r^{2} dr +\frac{q^{2} }{2r}
+\frac{1}{2} \int\frac{q^{2} }{r^{2} } dr. \label{eq30}
\end{equation}

Therefore, Eqs.~(\ref{eq15}) - (\ref{eq17}) in terms of the above
mass function $m(r)$ can be provided as
follows~\cite{Maurya2015a,Maurya2015b}:
\begin{equation}
-\frac{2 m}{r^{2} } \left[ \frac{ (1+r\nu ' )}{r} \right]
+\frac{\nu' }{r} +\frac{q^{2} (1+r\nu ' )}{r^{4} } +\frac{q^{2}
}{r^{4} } =\kappa p,\label{eq34}
\end{equation}

\begin{eqnarray}
\frac{-m' (r\nu ' +2)}{2r^{2} } -\frac{\, m}{2r^{2} } \left(
\frac{2r^{2} \nu '' +r^{2} \nu ^{\prime2} +\nu ' r -2}{r} \right)
\nonumber\\ + \left[\frac{2rqq' \nu ' -2q^{2} \nu ' +4qq' +(r^{2}
+q^{2} ) (2r\nu '' +r\nu ^{\prime2} +2\nu ' )}{4r^{3} }\right]
-\frac{2q^{2} }{r^{4} } =\kappa p, \label{eq35}
\end{eqnarray}

\begin{equation}
\frac{2m' }{r^{2} } -\frac{2qq' }{r^{3} } =\kappa \rho
\label{eq36}
\end{equation}

From Eqs. (\ref{eq34}) and (\ref{eq35}), the first order linear
differential equation for $m(r)$ in terms of $\nu(r)$ and electric
charge function $q(r)$ can be provided as:
\begin{equation}
m' +\frac{(2r^{2} \nu '' +r^{2} \nu ^{\prime2} -3\nu ' r -6)}{r
(r\nu ' +2)} m=\frac{r (2r\nu '' +r\nu ^{\prime2} -2\nu ' )}{2
(r\nu ' +2)} +f (r),\label{eq37}
\end{equation}
where
\begin{equation}
f(r)= \frac{q^{2} [ 2r^{2} \nu '' +r \nu ' (r\nu ' -4)-16]}{2
r^{2} (r\nu ' +2)} +\frac{qq'}{r},
\end{equation}
which gives the mass $m(r)$ as:
\begin{equation}
m (r)=e^{-\int g(r) dr } \left[ \int[ h (r) +f(r)] \left(
e^{\int\nolimits_{}^{}g (r) dr } \right) dr+A \right],
\label{eq38}
\end{equation}
where $ g(r)=\frac{(2r^{2} \nu '' +r^{2} \nu ^{\prime2} -3\nu ' r
-6)}{r (r\nu ' +2)}$ and $h(r)=\frac{r (2r\nu '' +r\nu ^{\prime2}
-2\nu ' )}{2 (r\nu ' +2)}.$

\section{A set of new class of solutions}
To find out a set of new class of solutions for electromagnetic
mass models let us now consider the following forms of the metric
potentials
\begin{equation}
\nu =2\log [ A+B\sqrt{1+br^{2} } ],\label{eq39a}
\end{equation}

\begin{equation}
\lambda=\log \left( \frac{1+a r^{2} }{1+br^{2} }
\right),\label{eq39b}
\end{equation}
where $A$ and $B$ are two positive constants with
\begin{equation}
B=\frac{1}{b} \sqrt{\frac{(a-b)}{K} }, \label{eq40}
\end{equation}
$a$ and $b$ being two real numbers.

Though the above forms of the metric potentials are chosen on the
{\it ad hoc} basis but later on one can see that these will lead
us to very interesting and physically valid solutions.

Thus the expressions for electromagnetic mass and electric charge
respectively can be provided as
\begin{equation}
m(r) =\frac{1}{2}r^{3} \left[ \frac{(a-b)}{1+ar^{2} }
+\frac{ar^{2} [C(r) - D(r)]}{F(r)} \right], \label{eq41}
\end{equation}

\begin{equation}
q(r) = Er^{2} =\sqrt{a}~r^{3}\sqrt{\left[ \frac{C(r) -
D(r)}{F(r)}\right]}, \label{eq42}
\end{equation}
where $C(r) =  a (B+bBr^{2} +A\sqrt{1+br^{2} })$, $D(r)= b
(A\sqrt{1+br^{2} } +2B+2Bbr^{2} )$ and $F(r) = 2(1+ar^2)^2
\sqrt{1+br^2} [A+B\sqrt{1+br^2}]$.

The expressions for fluid pressure and energy density are
respectively given by
\begin{equation}
8\pi p= \left[ \frac{-a^{2} r^{2}H(r)+2b[H(r)+2B(1+br^{2} )] +a
I_1}{F(r)} \right], \label{eq43}
\end{equation}

\begin{equation}
8\pi \rho = \left[ \frac{-6b H(r)+a^{2} r^{2} H(r)+a I_2}{F(r)}
\right], \label{eq44}
\end{equation}
where $H(r) = B(1+br^{2} )+A\sqrt{1+br^{2}}$, $I_1= A(-2+br^{2}
)\sqrt{1+br^{2} } +2B(-1+br^{2} +2b^{2} r^{4} )$ and
$I_2=A(6-br^{2} )\sqrt{1+br^{2} } +6B(1+br^{2} )$.

Therefore, the pressure and density gradients are
\begin{equation}
\frac{ dp}{dr} =\frac{2 r}{8\pi } \left[
\frac{p_1+p_2+p_3}{4(1+ar^2)^3 \sqrt{1+br^2} [A+B\sqrt{1+b
r^2}]^2} \right], \label{eq45}
\end{equation}

\begin{equation}
\frac{ d\rho }{dr} =-\frac{2A^{2} r}{8\pi } \left[
\frac{\rho_1+\rho_2+\rho_3}{4(1+ar^{2} )^{3} \sqrt{1+br^{2} } [
A+B\sqrt{1+b r^{2} } ]^{2} } \right],\label{eq46}
\end{equation}

where

$$p_1=4Ab^{2} B-a b [ 6A^{2} \sqrt{1+br^{2} } +16B^{2} (1+br^{2}
)^{3/2} +AB(22+15br^{2} )],$$

$$p_2=2a^{3} r^{2} [ A^{2} \sqrt{1+br^{2} } +B^{2} (1+br^{2}
)^{3/2} +2AB(1+br^{2} )],$$

$$p_3=a^{2} [ -2A^{2} (-3+br^{2} )\sqrt{1+br^{2} } +AB (12+2br^{2}
-7b^{2} r^{4} )-2B^{2} (-3+br^{2} +4b^{2} r^{4} ) \sqrt{1+br^{2} }
],$$

$$\rho_1=-b [ 22A^{2} \sqrt{1+br^{2} } +24B^{2} (1+br^{2} )^{3/2}
+AB(46+47br^{2} )],$$

$$\rho_2=2a^{2} r^{2} [ A^{2} \sqrt{1+br^{2} } +B^{2} (1+br^{2}
)^{3/2} +2AB(1+br^{2} )],$$

$$\rho_3=a[ -2A^{2} (-11+br^{2} )\sqrt{1+br^{2} } +22B^{2}
(1+br^{2} )^{3/2} +AB(44+42br^{2} -3b^{2} r^{4} )].$$

\subsection{Specific results at a Glance}
The metric (\ref{eq1}), with the metric potentials (\ref{eq39a})
and (\ref{eq39b}), describes the following special cases:

\subsubsection{$a=0$} If $a=0$ then corresponding
solution becomes the charge analogue of the
Schwarzschild~\cite{Schwarzschild1916} interior solution. In this
case of charged fluid sphere the metric potentials turn out to be
$e^{\nu} = (A+B\sqrt{1+br^{2}})^2$ and $e^{\lambda}=
(1+br^{2})^{-1}$.

\subsubsection{$a=2b$ with $A=0$} If one put $a=2b$ and $A=0$ in Eqs.~(30)
and (31) then corresponding solution becomes the charge analogue
of the Kohlar-Chao~\cite{Kohler1965} interior solution with the
metric potentials $e^{\nu} = B^2 (1+br^{2})$ and $e^{\lambda}=
(1+2br^{2})/(1+br^2)$.

\subsubsection{$a=b$} The case $a \neq b$ gives charged perfect fluid sphere
while the case $a=b$ implies flat spacetime with $B=0$, $e^{\nu} =
A^2$ and $e^{\lambda}=1$. As a consequence all the physical
parameters, viz.  mass, electric charge, corresponding pressure as
well as density become zero. This result is consistent with the
concept of electromagnetic mass model as proposed by
Lorentz~\cite{Lorentz1904}.

In the above analysis it would be curious, on the mathematical
point of view, to look at the possibility of replacement $a$ and
$b$ by $ka$ and $kb$ respectively where $k$ is a constant. On
making $k=0$ the metric turns out to be zero, whatever may be the
value of $a$ and $b$.

However, out of the above three cases we are interested for the
third sub-case {\it 4.1.3} which corresponds to the
electromagnetic mass model having a long historical background
with the structure of electron.

\section{Boundary conditions}
The arbitrary constants $A$, $B$ and $K$ can be obtained by using
the boundary conditions. For the above system of equations the
boundary conditions that applicable are as follows: the pressure
$p=0$ at $r=R$, where $r=R$ is the outer boundary of the fluid
sphere. Actually, the interior metric (\ref{eq1}) should join
smoothly at the surface of spheres $(r=R)$ to the exterior
Reissner-Nordstr{\"o}m metric whose mass is $m(r=R)=M$, a constant
\cite{Misner1964}, given by
\begin{equation}
ds^{2} =-\left( 1-\frac{2M}{r} +\frac{Q^{2} }{r^{2} } \right)
^{-1} dr^{2} -r^{2} (d\theta ^{2} +\sin ^{2} \theta d\phi ^{2}
)+\left( 1-\frac{2M}{r} +\frac{Q^{2} }{r^{2} } \right) dt^{2}.
\label{eq47}
\end{equation}
This requires the continuity of $e^{\lambda } $, $ e^{\nu } $ and
$Q$ across the boundary $r=R$, so that
\begin{equation}
e^{-\lambda (R)} =1-\frac{2M}{R} +\frac{Q^{2} }{R^{2} },
\label{eq48}
\end{equation}

\begin{equation}
e^{\nu (R)} = 1-\frac{2M}{R} +\frac{Q^{2} }{R^{2} }, \label{eq49}
\end{equation}

\begin{equation}
q(R)=Q, \label{eq50}
\end{equation}

\begin{equation}
p_{(r=R)} =0.\label{eq51}
\end{equation}

The pressure is zero on the boundary $r=R $ and hence we obtain
\begin{equation}
\frac{B}{A} =\frac{ (a-b) (2+aR^{2} )}{[ 6b-a^{2} R^{2} -a
(2-4bR^{2} )] \sqrt{1+bR^{2} } }. \label{eq52}
\end{equation}

Again, at the boundary $e^{-\lambda (R)} =e^{\nu (R)}$, which
gives
\begin{equation}
A=\frac{(6 b-a^{2} R^{2} -2a+4abR^{2} ) \sqrt{1+bR^{2} }
}{(4b+3abR^{2} ) \sqrt{1+aR^{2} } }. \label{eq53}
\end{equation}

Also from Eqs. (\ref{eq52}) and (\ref{eq53}) one gets
\begin{equation}
B=\frac{(a-b) (2+aR^{2} )}{(4b+3abR^{2} ) \sqrt{1+aR^{2} } }.
\label{eq54}
\end{equation}

For the third constant $K$, we use the Eqs.~(\ref{eq40}) and
(\ref{eq54}), which provides the required expression as
\begin{equation}
K=\frac{ (4b+3abR^{2} )^{2} (1+aR^{2} )}{b^{2} (a-b) (2+a R^{2}
)^{2} }. \label{eq55}
\end{equation}

\section{Physical acceptability conditions for the isotropic stellar models}
In this Sec. 6 we have critically verified our models by
performing mathematical analysis and plotting several figures for
some of the compact star candidates. All these indicate that the
results are fantastically overcome all the barrier of the physical
tests.

\subsection{Regularity and Reality Conditions}

\subsubsection{Case 1}

It is expected that the solution should be free from physical and
geometrical singularities i.e. the pressure and energy density at
the centre should be finite and metric potentials $e^{\lambda
(r)}$ and $e^{\nu (r)}$ should have non-zero positive values in
the range $0 \le r \le R$. We observe that at the centre Eqs.~(30)
and (31) gives $e^{\lambda (0)} =1$ and $e^{\nu (0)} =(A+B)^{2} $.
These results suggest that the metric potentials are positive and
finite at the centre. These features can be found explicitly from
Fig.~1.

\begin{figure}[h]
\centering
\includegraphics[scale=.6]{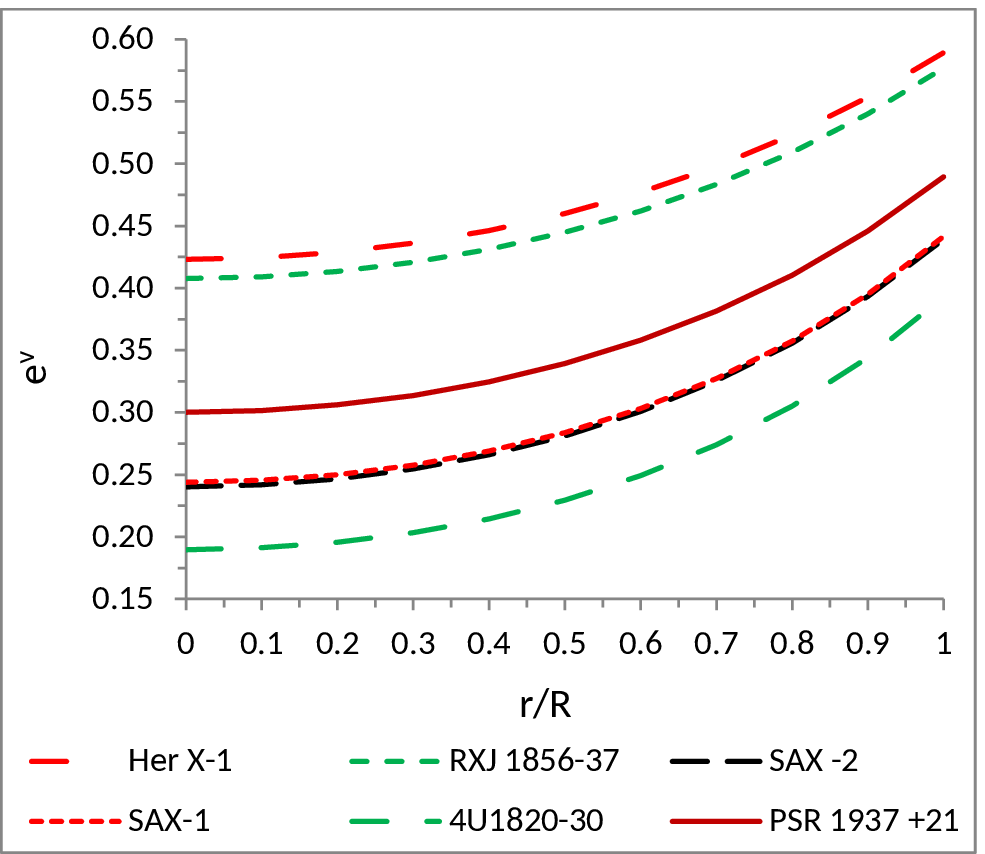}
\includegraphics[scale=.6]{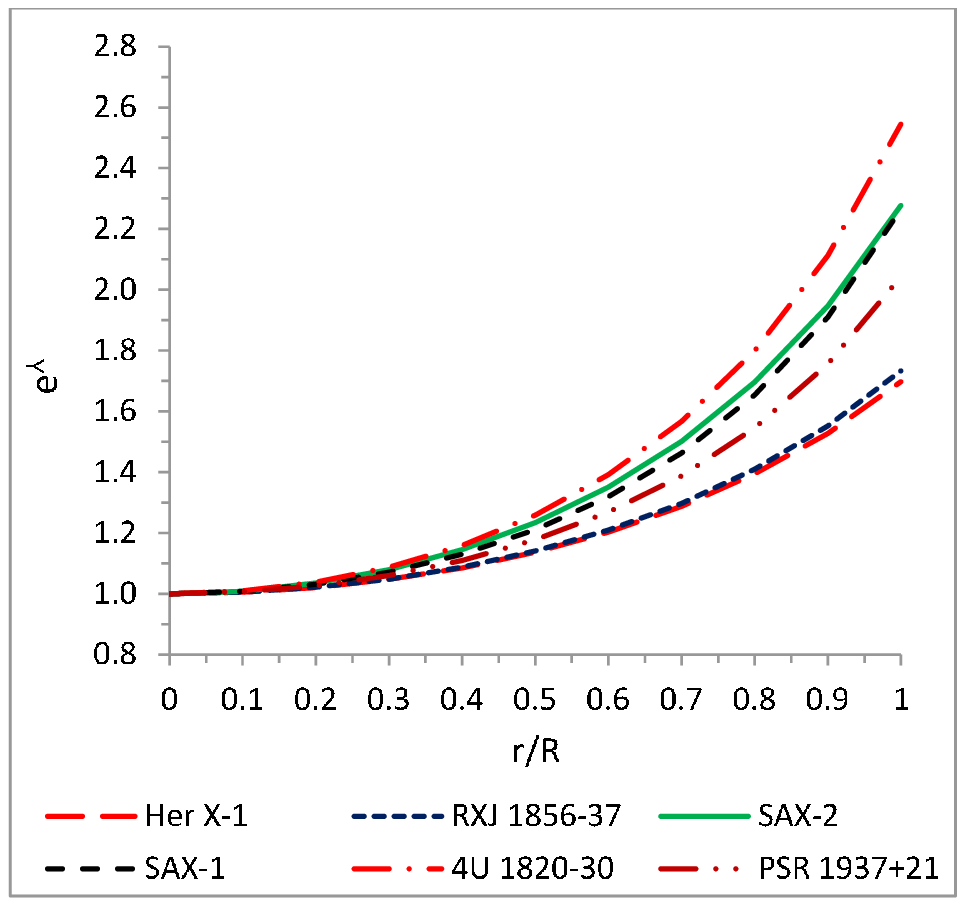}
\caption{The behavior of metric potentials $\nu$ and $\lambda$
with respect to radial coordinate $r/R$ }
\end{figure}

\subsubsection{Case 2}
For any physical valid solutions the density $\rho$ and pressure
$p$ should be positive inside the star. Also the pressure must
vanish on the boundary of the fluid sphere $r=R$. The other
physical conditions to be maintained are as follows:

(1) $\left( dp/dr\right) _{r=0} =0 $ and $\left( d^{2} p/dr^{2}
\right) _{r=0} <0$ so that pressure gradient $dp/dr $ is negative
for $0 \le r \le R$.

(2) $\left( d\rho /dr\right) _{r=0} =0 $ and $\left( d^{2} \rho
/dr^{2} \right) _{r=0} <0 $ so that density gradient $d\rho /dr $
is negative for $0 \le r \le R$.

The above two conditions (1) and (2) imply that the pressure and
density should be maximum at the centre and they should
monotonically decrease towards the surface. All these are evident
from Fig.~2.

\begin{figure}[h]
\centering
\includegraphics[scale=.6]{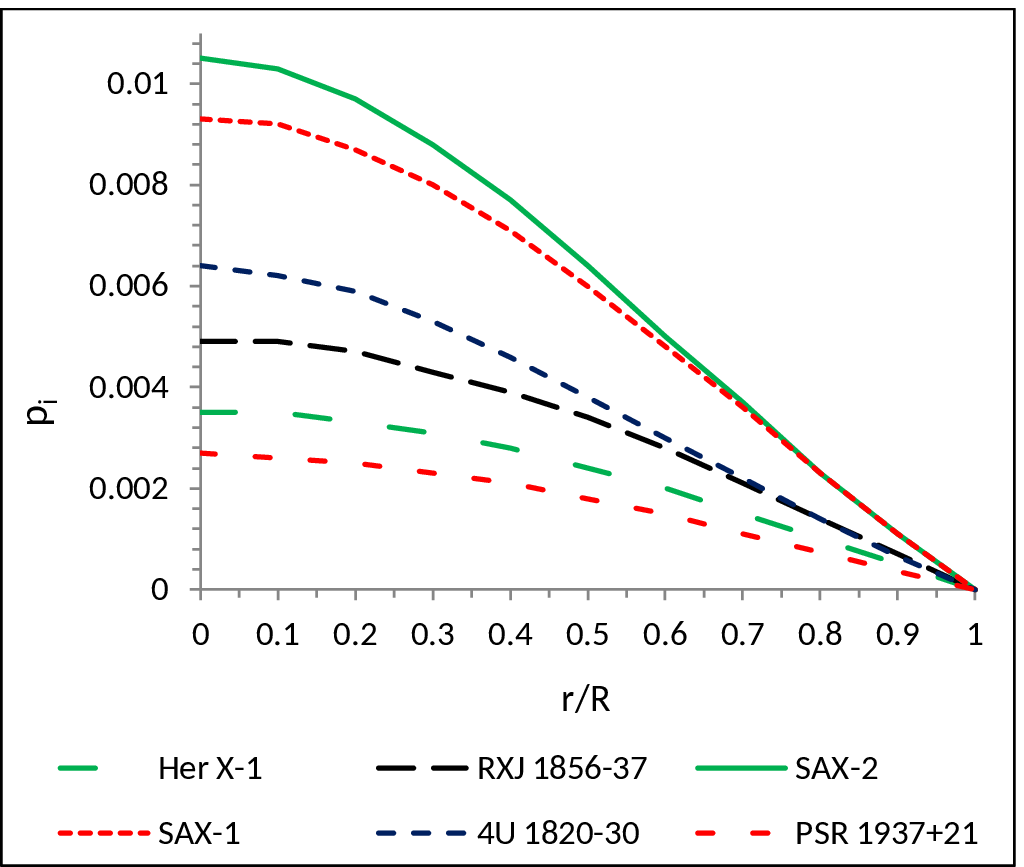}
\includegraphics[scale=.6]{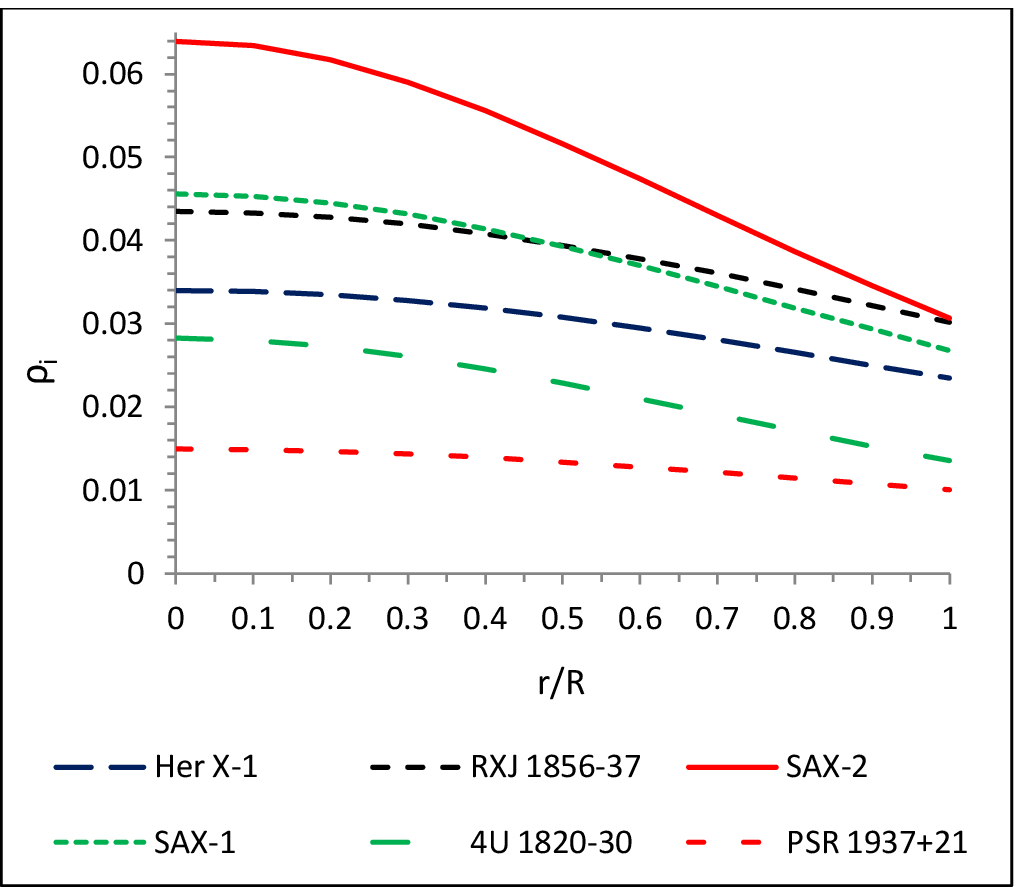}
\caption{The behavior of fluid pressure $p$ and energy density
$\rho$ with respect to radial coordinate $r/R$, where $p_i=\kappa
p$, $\rho_i= \kappa \rho$. }
\end{figure}

\subsection{Causality and Well Behaved Conditions}
Inside the fluid sphere the speed of sound should be less than the
speed of light i.e. $0\leq \left(\frac{dp}{d\rho} \right) <1 $,
which can be observed in Fig.~3. We observe from this figure that
the velocity of sound monotonically is decreasing away from the
centre \cite{Canuto1973}.

\begin{figure}[h]
\centering
\includegraphics[scale=.6]{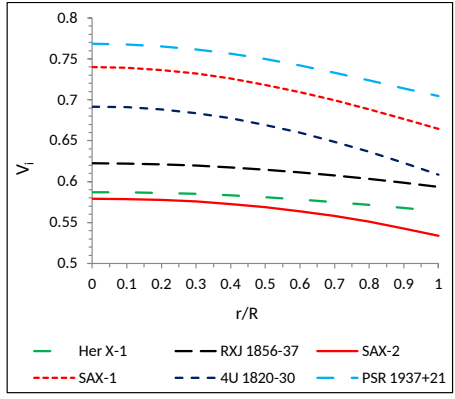}
\caption{The behavior of sound speed $v$ with respect to radial
coordinate $r/R$ }
\end{figure}

\subsection{Energy Conditions}
It is, in general, argued that a physically reasonable
energy-momentum tensor which represents an isotropic charged fluid
sphere composed of matter must satisfy the following energy
conditions:
\begin{enumerate}
\item null energy condition (NEC): $\rho+\frac{E^2}{4\pi} \geq 0$
\item weak energy condition (WEC): $\rho-p+\frac{E^2}{4\pi} \geq 0$
\item strong energy condition (SEC): $\rho-3p+\frac{E^2}{4\pi} \geq 0$
\end{enumerate}

The behaviour of these energy conditions are shown in Fig.~4. This
figure clearly indicates that all the energy conditions in our
model are satisfied throughout the interior region of the
spherical distribution.

\begin{figure}[h]
\centering
\includegraphics[scale=.6]{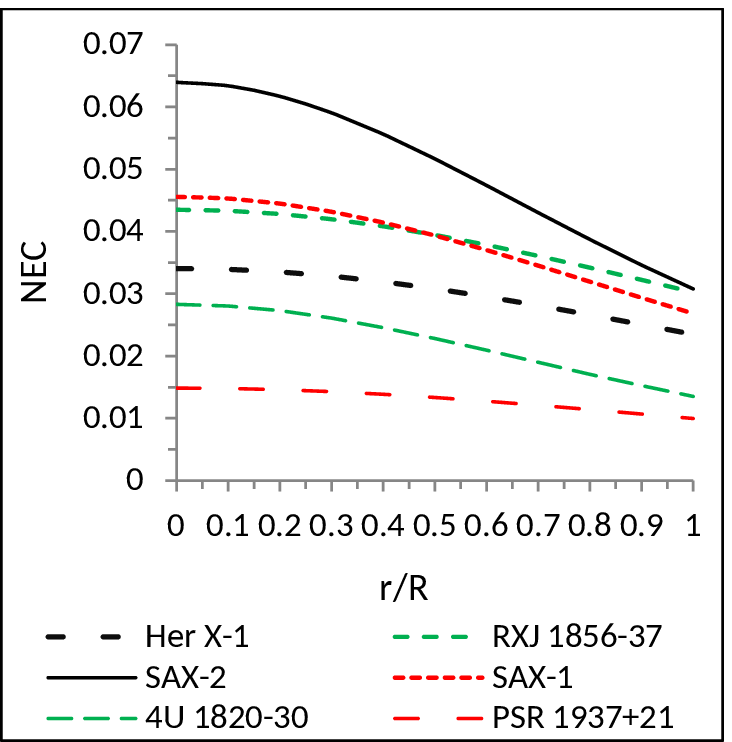}
\includegraphics[scale=.6]{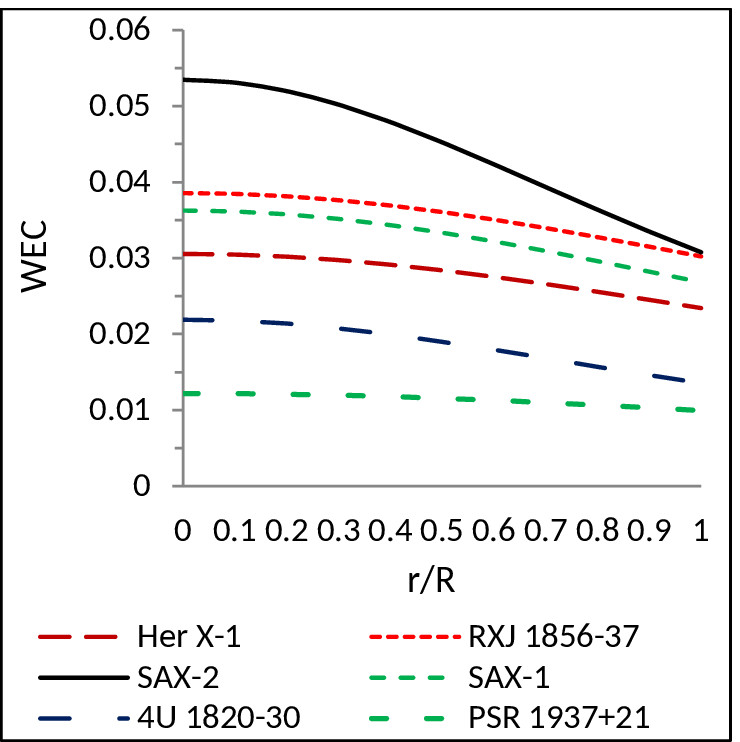}
\includegraphics[scale=.6]{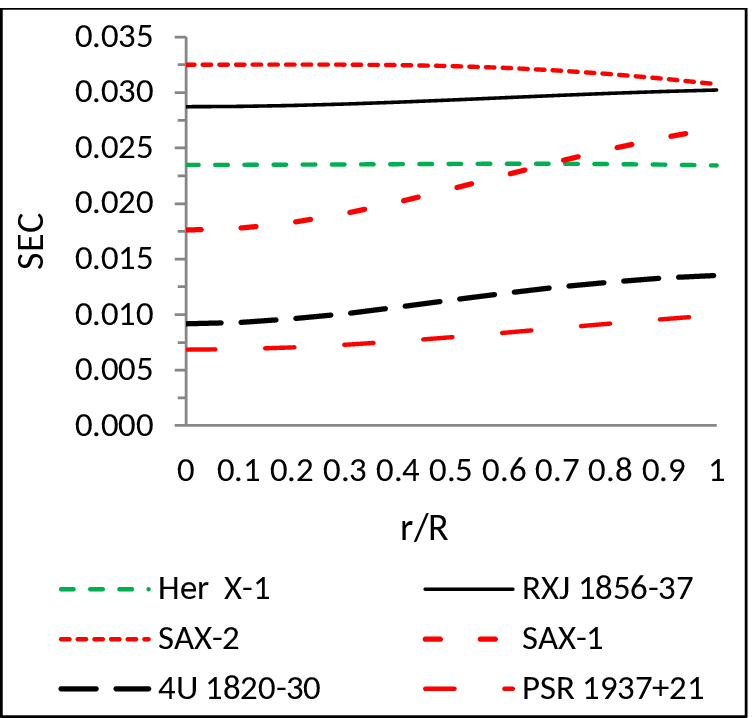}
\caption{The behavior of the energy conditions with respect to
fractional radius $r/R$}
\end{figure}

\subsection{Stability Conditions}

\subsubsection{Tolman-Oppenheimer-Volkoff equation} The generalized
Tolman-Oppenheimer-Volkoff (TOV) equation can be provided as
\begin{equation}
-\frac{M_{G} (\rho +p_{r} )}{r^{2} } e^{\frac{\lambda -\nu }{2} }
-\frac{dp}{dr} +\sigma \frac{q}{r^{2} } e^{\frac{\lambda }{2} }
=0, \label{eq56}
\end{equation}
where $M_{G} $ is the effective gravitational mass given by
\begin{equation}
M_{G} (r)=\frac{1}{2} r^{2} e^{\frac{\nu -\lambda }{2} } \nu'.
\label{eq57}
\end{equation}
This Eq. ({\ref{eq56}) describes the equilibrium condition for a
charged perfect fluid subject to the gravitational $(F_{g})$,
hydrostatic $(F_{h})$ and electric $(F_{e})$. In summary, we can
write it as
\begin{equation}
F_{g} + F_{h} + F_{e} = 0,\label{eq58}
\end{equation}
where
\begin{equation}
F_{g} =-\frac{1}{2} (\rho +p)\nu' = -\frac{ rbB}{8\pi} \left[
\frac{2[A(a-b)\sqrt{1+br^2} + B a(1+b r^2)]}{(1+ar^2)^2 (1+br^2)
[A+B \sqrt{1+br^2}]^2}\right], \label{eq59}
\end{equation}

\begin{equation}
F_{h} =-\frac{ dp}{dr} =-\frac{2 r}{8\pi } \left[
\frac{p1+p2+p3}{4(1+ar^{2} )^{3} \sqrt{1+br^{2} } [ A+B\sqrt{1+b
r^{2} } ]^{2} } \right],\label{eq60}
\end{equation}

\begin{equation}
F_{e} =\sigma \frac{q}{r^2}e^{\lambda/2}=\frac{a r}{4\pi } \left[
\frac{F_{e1} +F_{e2} +F_{e3} }{4(1+ar^{2} )^{3} \sqrt{1+br^{2} } [
A+B\sqrt{1+b r^{2} } ]^{2} } \right],\label{eq61}
\end{equation}

$$p1=4Ab^{2} B-a b [ 6A^{2} \sqrt{1+br^{2} } +16B^{2} (1+br^{2}
)^{3/2} +AB(22+15br^{2} )],$$

$$p2=2a^{3} r^{2} [ A^{2} \sqrt{1+br^{2} } +B^{2} (1+br^{2}
)^{3/2} +2AB(1+br^{2} )],$$

$$p3=a^{2} [ -2A^{2} (-3+br^{2} )\sqrt{1+br^{2} } +AB (12+2br^{2}
-7b^{2} r^{4} ) -2B^{2} (-3+br^{2} +4b^{2} r^{4} ) \sqrt{1+br^{2}
} ],$$

$$F_{e1} =-b [ 6A^{2} \sqrt{1+br^{2} } +12B^{2} (1+br^{2} )^{3/2}
+AB (18+19br^{2} )],$$

$$F_{e2} =2a^{2} r^{2} [ A^{2} \sqrt{1+br^{2} } +B^{2} (1+br^{2}
)^{3/2} +2AB(1+br^{2} )],$$

$$F_{e3} =a [ -2A^{2} (-3+br^{2} )\sqrt{1+br^{2} } +AB (12+6br^{2}
-7b^{2} r^{4} ) +2B^{2} (3+br^{2} -2b^{2} r^{4} )\sqrt{1+br^{2} }
].$$

We have shown the plots for TOV equation in Fig.~5 for different
compact strange stars. From the figures it is observed that the
system is in static equilibrium under four different forces, e.g.
gravitational, hydrostatic, electric and anisotropic to attain
overall equilibrium.

\begin{figure}[h]
\centering
\includegraphics[scale=.6]{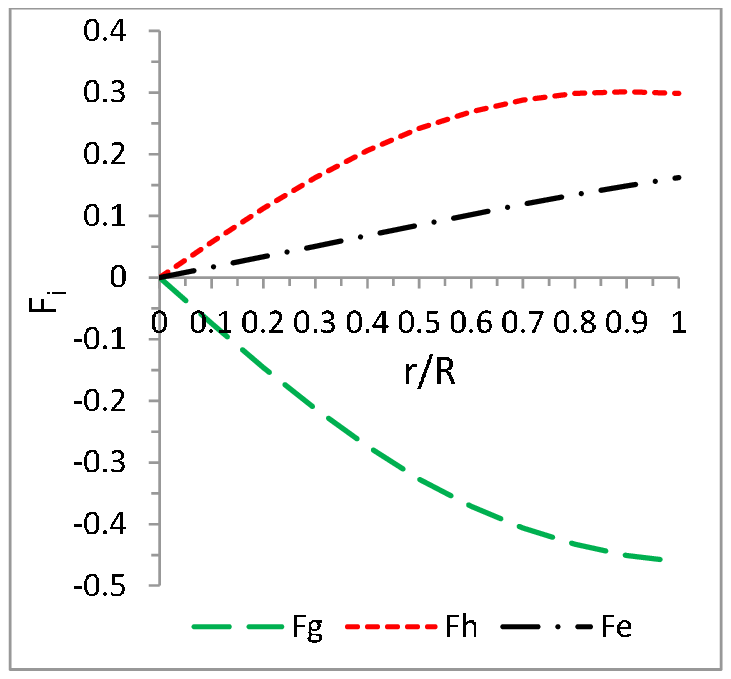}
\includegraphics[scale=.6]{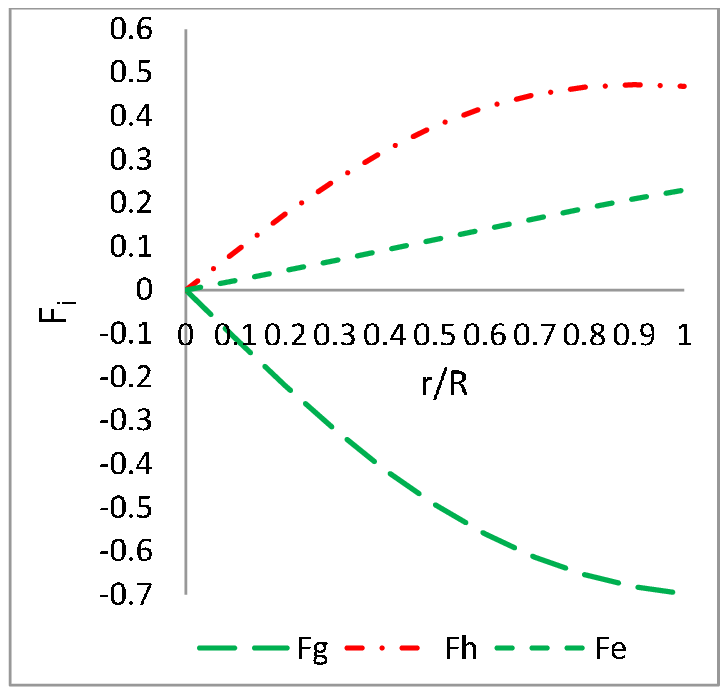}
\includegraphics[scale=.6]{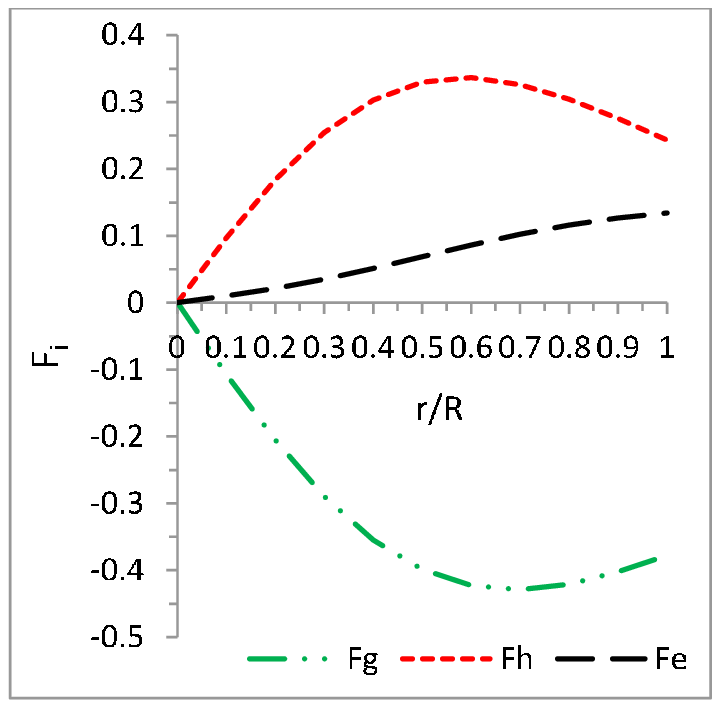}
\includegraphics[scale=.6]{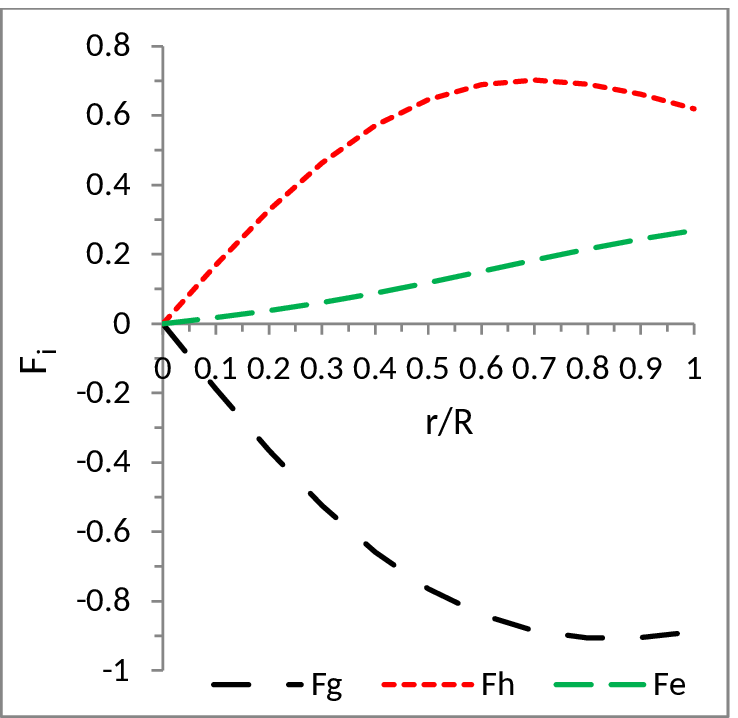}
\includegraphics[scale=.6]{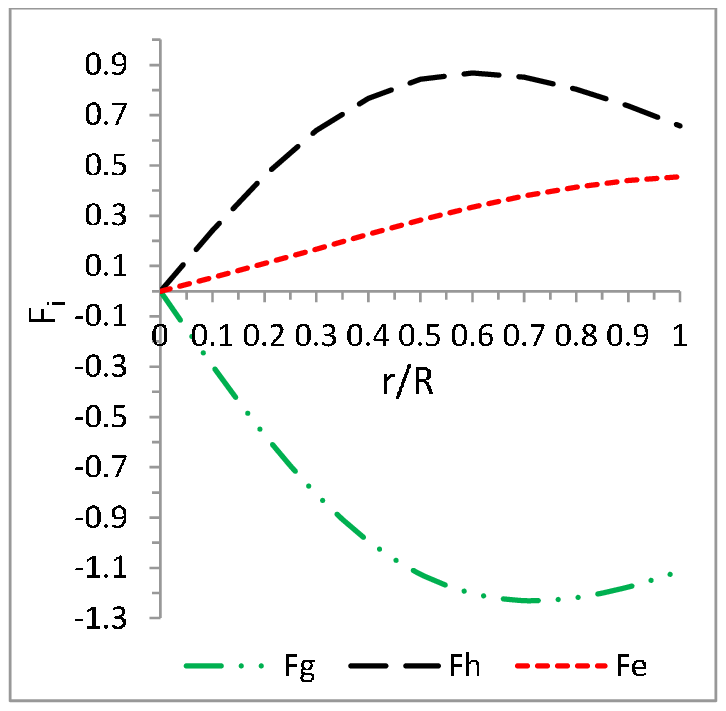}
\includegraphics[scale=.6]{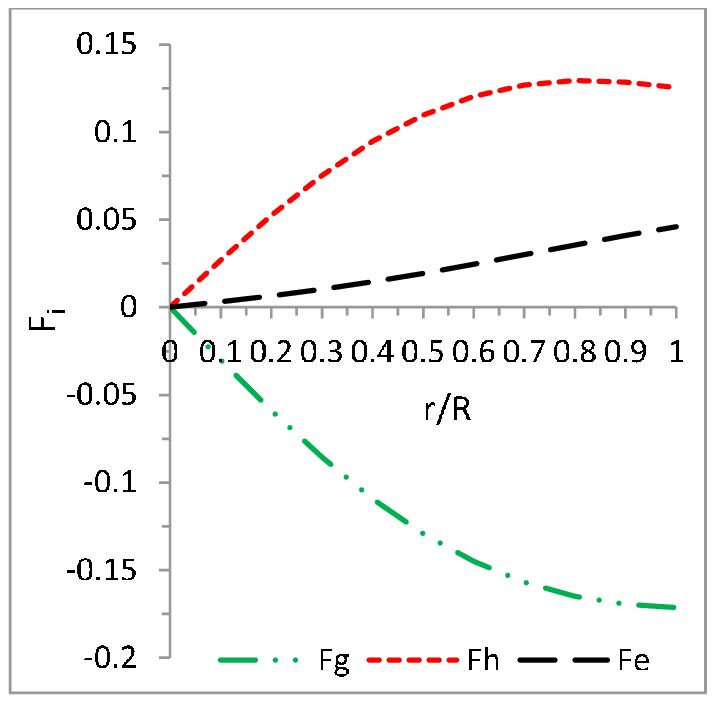}
\caption{The behavior of forces for the compact stars (i)
$Her~X-1$ (Top Left), (ii) $RX~J~1856-37$ (Top Middle) and (iii)
$4U~1820-30$ (Top Right) (iv) $SAX~J1808.4-3658 (SS1)$ (Bottom
Left), (v) $SAX~J1808.4-3658 (SS2)$ (Bottom Middle) and (vi)
$PSR~1937+21$ (Bottom Right) with respect to radial coordinate
$r/R$ }
\end{figure}

\subsubsection{Electric charge contain} In the present work the
expression for electric charge can be given by Eq.~(\ref{eq42})
and following the work of Maurya et al.~\cite{Maurya2015a} we can
figure out that the charge on the boundary is $1.15295 \times
10^{20}$~C and at the center it is zero. The charge profile has
been shown in the Fig. 6 for different compact stars which starts
from a minimum value at the centre and acquires the maximum value
at the boundary. This feature is also evident from the Table 1 and
compatible with the result of Ray et al.~\cite{Ray2003} where they
studied the effect of electric charge in compact stars and found
the upper bound as $\sim 10^{20}$ Coulomb.

\begin{figure}[h]
\centering
\includegraphics[scale=.6]{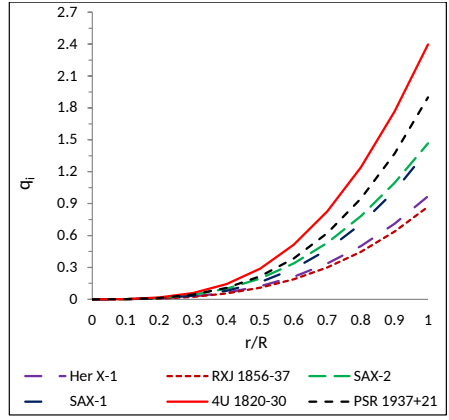}
\caption{Behavior of electric charge $q$ with respect to radial
coordinate $r/R$ }
\end{figure}

\begin{table}[h] \centering \caption{The profile of electric charge for different compact stars}\label{tbl-4}
\begin{tabular}{@{}lrrrrrrrrrrrrr@{}} \hline

$r/a$ & Her. X-1 &  RXJ 1856-37 & SAX-2 &  SAX-1 &  4U 1820-30 &
PSR 1937+21\\ \hline

0.0 & 0 &  0  & 0 &  0 &  0 &  0 \\

1.0 & 0.9686 & 0.8706 & 1.4664 & 1.4056 & 2.3974 & 1.8996\\ \hline
\end{tabular}
\end{table}

\subsubsection{Effective mass-radius relation}
Buchdahl \cite{Buchdahl1959} has proposed an absolute constraint
on the maximally allowable mass-to-radius ratio $(M/R)$ for static
spherically symmetric isotropic fluid spheres which amounts
$2M/R\leq 8/9 $. On the other hand, B{\"o}hmer and Harko
\cite{Boehmer2007} proved that for a compact charged fluid sphere
there is a lower bound for the mass-radius ratio
\begin{equation}
\frac{3Q^{2} }{2R^{2} } \frac{\left( 1+\frac{Q^{2} }{18R^{2} }
\right) }{\left( 1+\frac{Q^{2} }{12R^{2} } \right) } \leq
\frac{2M}{R}, \label{eq62}
\end{equation}
for the constraint $Q < M$.

This upper bound of the mass for charged fluid sphere was
generalized by Andreasson \cite{Andreasson2009} who proved that
\begin{equation}
\sqrt{M} \leq \frac{\sqrt{R} }{3} +\sqrt{\frac{R}{9} +\frac{Q^{2}
}{3R} }. \label{eq63}
\end{equation}

In the present model, we find the effective gravitational mass as
\begin{equation}
M_{eff} =4\pi \int\nolimits_{0}^{R}\left( \rho +\frac{E^{2} }{8\pi
} \right) r^{2} dr= \frac{1}{2} R(1-e^{-\lambda (R)} ) =
\frac{1}{2} R \left[ \frac{(a-b) R^{2} }{(1+a R^{2} )} \right].
\label{eq64}
\end{equation}

In terms of the compactness factor $u=M_{eff}/R$ we now define the
surface red-shift $Z_{s}$ as
\begin{equation}
Z_{s} =(1-2u)^{-\frac{1}{2} } -1=e^{\frac{1}{2} \lambda (R)} -1=
\sqrt{\frac{(1+aR^{2} )}{(1+bR^{2} )} } -1. \label{eq67}
\end{equation}

We have demonstrated the behavior of surface redshift $Z_s$ with
respect to radial coordinate $r/R$ in Fig. 7 which shows the
desirable features~\cite{Maurya2015b}.

\begin{figure}[h]
\centering
\includegraphics[scale=.6]{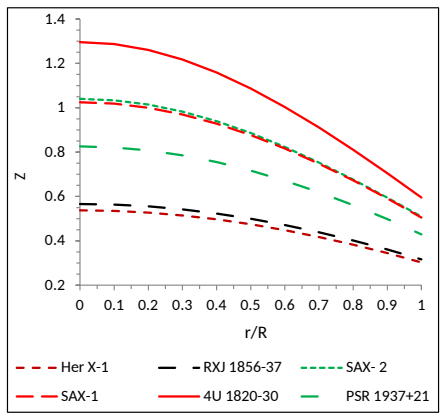}
\caption{The behavior of surface redshift $Z_s$ with respect to
radial coordinate $r/R$ }
\end{figure}

\section{Some special features of the models}

\subsection{Stellar structure}
In the previous Sec. 6, we have discussed several properties of
our solutions in terms of various physical parameters based on
some of the compact stars. In this Sub-section we provide two
Tables 2 and 3 where we have figured out some other physical
parameters as well as some constants of our models. Actually, the
values of Table 2 have been used in Table 3 to find out the energy
densities and pressure for different strange star candidates. It
is worthwhile to mention that the densities $\sim
10^{15}$~$gm/cm^{-3}$ and pressure $\sim 10^{35}$~$dyne/cm^{-2}$
are in very good agreement with the observational data of the
compact stars, specially $Her~X~1$
\cite{Ruderman1972,Maurya2015c}.

\begin{table}[h]
\centering \caption{Values of the model parameters for different
strange stars}\label{tbl-2}
\begin{tabular}{@{}lrrrrrrrrrrrrr@{}} \hline
Strange star &$M$ & $R$ & $M/R$ & $a$ & $b$ & $A$ &$K$  \\
candidates & $(M_\odot)$ & $(Km)$ \\ \hline

$Her~X-1$  & 0.9800 & 6.70 & 0.2160 &0.00535 &-0.00600 & 1.4566 &
485.1733\\

$RX~J~1856-37$ & 0.9000 & 6.00 & 0.2220 & 0.00650 & -0.00800 &
1.4137  &377.0041\\

$SAX~J1808.4-3658 (SS1)$  & 1.4351 & 7.07 & 0.2990 & 0.00719 &
-0.00800 & 1.2506 & 414.5739\\

$SAX~J1808.4-3658 (SS2)$ & 1.3235 & 6.35 & 0.3071 & 0.01320 &
-0.00811 & 1.4506  & 351.2430\\

$PSR~1937+21$ & 2.0830 & 11.40 & 0.2692 & 0.00200 & -0.00295 &
1.2546 & 1138.6000\\

$4U~~1820-30$ & 2.2457 & 9.95 & 0.3325 & 0.00542 & -0.00400 &
1.2946 & 797.8075 \\
\hline
\end{tabular}
\end{table}

\vspace{1.0cm}

\begin{table}[h]
\centering \caption{Energy densities and pressure for different
strange star candidates \newline for the above parameter values of
Table 1}\label{tbl-3}
\begin{tabular}{@{}lrrrrrrrrr@{}}
\hline

Strange star & Central energy density & Surface energy density &
Central pressure \\

candidates & ($gm/cm^{-3}$) & ($gm/cm^{-3}$) & ($dyne/cm^{-2}$)\\
\hline

$Her~X-1$      &  $1.8285 \times 10^{15}$ & $1.2590 \times
10^{15}$ & $1.7018 \times 10^{35}$\\

$RXJ~1856-37$  &  $2.3360 \times 10^{15}$  & $1.6223 \times
10^{15}$ & $2.3810 \times 10^{35}$\\

$SAX~J1808.4-3658 (SS1)$  & $2.4471 \times 10^{15}$ & $1.4404
\times 10^{15}$ & $4.5051 \times 10^{35}$\\

$SAX~J1808.4-3658 (SS2)$  & $3.4330 \times 10^{15}$ & $1.6506
\times 10^{15}$ & $5.0610 \times 10^{35}$\\

$PSR~1937+21$  &  $1.5176 \times 10^{15}$ & $7.2620 \times
10^{14}$ & $3.0741 \times 10^{35}$\\

$4U~1820-30$ & $7.9745 \times 10^{15}$ & $5.3441 \times 10^{14}$ &
$1.2871 \times 10^{35}$\\ \hline

\end{tabular}
\end{table}

\subsection{Electronic structure}
Let us now come down from macro-scale of the stellar structure to
the micro-scale of the structure of the electron. Here we have
performed a comparative study of the values of the physical
parameters of electron between the data of the present work and
that from the work of Gautreau \cite{Gautreau1985} as shown in
Table 4. Here also one can observe that the data of both works are
exactly correspond to each other at least as far as order of
magnitude are concerned. In is to note that the density of
electron from our model turns out to be $8.541 \times
10^{10}~gm/cm^3$ which seems to closer to the actual value of the
density of electron.

\begin{table}[h] \centering \caption{A comparative study of
the values of the physical parameters of electron}\label{tbl-4}
\begin{tabular}{@{}lrrrrrrrrrrrrr@{}} \hline
Physical parameter & In the present paper & Data from Gautreau
\cite{Gautreau1985}\\ \hline

Mass  & $6.6772 \times 10^{-56}$~cm & $6.67 \times 10^{-56}$~cm\\

Radius & $2.82 \times 10^{-13}$~cm & $2.82 \times 10^{-13}$~cm\\

Charge  & $1.95 \times 10^{-34}$~cm & $1.38 \times 10^{-34}$~cm\\

\hline
\end{tabular}
\end{table}

For specific numerical values of the constant $\kappa=8\pi G/c^4$
and other physical parameters we have used the data $G=6.67 \times
10^{-8} cm^3/gs^{-2}$ and $c=2.997 \times 10^{10} cm/s$ in the
calculations of Tables 3 and 4.

\section{Conclusion}

Our sole aim in the present letter was to investigate nature of
class 1 metric. For this purpose we have considered matter-energy
distribution under the framework of Einstein-Maxwell spacetime. At
first we developed an algorithm which has a general nature and
thus can be reduced to three special cases, viz. (i) charge
analogue of the Kohler-Chao~\cite{Kohler1965} solution, (ii)
charge analogue of the Schwartzschild~\cite{Schwarzschild1916}
solution (i.e. the Reissner-Nordstro{\"o}m solution), and (iii)
the Lorentz~\cite{Lorentz1904} solution of electromagnetic mass
model.

By considering the third case of the Lorentz solution of
electromagnetic mass model we have studied its properties through
the following two basic physical testing, such as (i) regularity
and reality conditions, and (ii) causality and well behaved
conditions. Moreover, some other essential testing also have been
performed, viz. (i) energy conditions, and (ii) stability
conditions. In the case of energy conditions we have seen that the
isotropic charged fluid sphere composed of matter satisfy the (i)
null energy condition $(\rho+\frac{E^2}{4\pi}) \geq 0$, (ii) weak
energy condition $(\rho-p+\frac{E^2}{4\pi}) \geq 0$, and (iii)
strong energy condition $(\rho-3p+\frac{E^2}{4\pi}) \geq 0$
(Fig.~4). On the other hand, in connection to stability conditions
we critically have discussed the Tolman-Oppenheimer-Volkoff
equation, electric charge contain, effective mass-radius relation
of the charged spherical distribution. Here also we find that the
results are in favour of the physical requirements (Figs. 5 - 8).

As some special features of the models we have presented here
two-level applications in the following fields: (i) stellar
structure, and (ii) electronic structure. The behavior of the
compact stars (i) $Her~X-1$, (ii) $RX~J~1856-37$, (iii)
$4U~1820-30$, (iv) $SAX~J1808.4-3658 (SS1)$, (v) $SAX~J1808.4-3658
(SS2)$, and (vi) $PSR~1937+21$ have been demonstrated through two
Tables 2 and 3 which are quite satisfactory. Another application
of the models have been done in the case of the electron. This is
shown in the Table 4 where one can notice that the model data
resembles with the observational data of the electron.

\section*{Acknowledgments}
SKM acknowledges support from the authority of University of
Nizwa, Nizwa, Sultanate of Oman. Also SR is thankful to the
authority of Inter-University Center for Astronomy and
Astrophysics, Pune, India for providing Associateship under which
a part of this work was carried out.

\end{document}